\documentclass[12pt,twoside]{article} 
\usepackage[latin1]{inputenc}
\usepackage{graphicx}
\usepackage{latexsym}
\usepackage{amssymb}
\usepackage{epsfig}
\usepackage{hyperref}
\usepackage{amsmath}
\usepackage{natbib}

\newcommand{\Var}{\mbox{Var}}
\def\eps{\varepsilon}

\newcommand{\ds}{\displaystyle}

\def\en{\mathbb{N}}
\def\zet{\mathbb{Z}}
\def\er{\mathbb{R}}
\def\ge{\mathbb{G}}
\def\ka{\mathbb{K}}
\def\be{\mathbb{B}}

\def\e{\varepsilon}
\def\he{\hat{\varepsilon}}

\def\sjn{\sum_{j=1}^n}
\def\op{o_P(\frac 1{\sqrt n})}

\def\beq{\begin{eqnarray*}}
\def\eeq{\end{eqnarray*}}

\def\bK{\bar{K}}

\def\nn{\nonumber}

\begin{document}

\title{\bf A note on nonparametric testing for Gaussian innovations in AR-ARCH models
}

\author{Natalie Neumeyer and Leonie Selk\footnote{corresponding author, e-mail: leonie.selk@math.uni-hamburg.de} \\
University of Hamburg, Department of Mathematics \\
Bundesstrasse 55, 20146 Hamburg, Germany\\
}

\date{April 18, 2012}

\maketitle

\newtheorem{theo}{Theorem}[section]
\newtheorem{lemma}[theo]{Lemma}
\newtheorem{cor}[theo]{Corollary}
\newtheorem{rem}[theo]{Remark}
\newtheorem{prop}[theo]{Proposition}
\newtheorem{defin}[theo]{Definition}
\newtheorem{example}[theo]{Example}

\begin{abstract}In this paper we consider autoregressive models with conditional autoregressive variance, including the case of homoscedastic AR-models and the case of ARCH models. 
Our aim is to test the hypothesis of normality for the innovations in a completely nonparametric way, i.\,e.\ without imposing parametric assumptions on the conditional mean and volatility functions. To this end the CramÚr-von Mises test based on  the empirical distribution function of nonparametrically estimated residuals is shown to be asymptotically distribution-free. We demonstrate its good performance for finite sample sizes in a simulation study. 
\end{abstract}



\newpage

\section{Introduction}
\def\theequation{1.\arabic{equation}}
\setcounter{equation}{0}

Nonlinear AR-ARCH models, i.\,e.\ models with an autoregressive conditional mean function and an autoregressive conditional variance function which are both not assumed linear, have become increasingly popular. They are also called CHARN (conditional heteroscedastic autoregressive nonlinear) models. In this paper we assume an AR-ARCH model of order one. Our aim is to test for Gaussian distribution of the innovations, which constitutes a typical assumption in the modelling of time series data. Under the normality assumption, asymptotic results often simplify. For instance, then no fourth moments appear in the asymptotic variance matrix of the empirical autocovariances of linear processes, see e.\,g. Brockwell \& Davis (2006), Proposition 7.3.3.
Further, the lack-of-fit test for ARCH models by Horvßth, Kokozka \& TeyssiÞre (2004) strongly depends on the assumption of Gaussian innovations.
Araveeporn (2011) uses the assumption of Gaussian innovations in order to estimate the conditional mean and volatility functions.  
Furthermore, estimation of conditional quantiles is of great importance in the context of financial time series. Starting from a nonparametric AR-ARCH model, however, the quantiles of the innovation distribution need to be estimated, or better be known (see Franke, Krei▀ \& Mammen (2009), section 4). Moreover, under Gaussianity of the innovations one can derive asymptotically distribution-free versions of other specification tests; see the discussion below. 

We suggest a completely nonparametric test, which does not assume any parametric assumption on either mean or volatility function, but applies kernel estimators for those functions (see Doukhan \& GhindÞs (1983), Robinson (1983), Masry \& Tj$\o$stheim (1995), Hõrdle \& Tsybakov (1997), among others, for estimation procedures in nonparametric AR-ARCH models). Relatedly, in a homoscedastic nonparametric AR model M³ller, Schick \& Wefelmeyer (2009), who prove an asymptotic expansion of the empirical process of estimated innovations, mention the possibility to use their result for testing goodness-of-fit of the innovation distributions. They do not present the asymptotic distribution of the test statistics, nor finite sample properties. On the other hand, goodness-of-fit tests for the innovation distribution in parametric time series are suggested by Koul \& Ling (2006) in AR-ARCH models and by Klar, Lindner \& Meintanis (2011) for GARCH models, for instance. Ducharme \& Lafaye de Micheaux (2004) test for normality of the innovations in standard ARMA models.

Our test statistic for Gaussianity of the innovations is  the CramÚr-von Mises distance of a weighted empirical distribution function of estimated innovations and the standard normal distribution. Though the mean and volatility functions are not specified in any way, the test statistic is shown to be asymptotically distribution-free. The test and its asymptotic distribution are presented in section \ref{section-AR-ARCH}. 
 We treat the special cases of nonparametric AR and nonparametric ARCH models in detail in sections \ref{section-AR} and \ref{section-ARCH}. 
 In a small simulation study we demonstrate the good performance of the test for moderate sample sizes in section \ref{section-simus}.  We further discuss briefly how the test can be generalized to AR-ARCH models of higher order or to models with additional covariates. 

As already mentioned, under Gaussianity of the innovations other testing procedures based on the residual empirical process will be asymptotically distribution-free. Then bootstrap procedures (for which asymptotic validity often is not investigated rigorously in the literature) can be avoided. As example for an asymptotically distribution-free specification test under the normality assumption we present a lack-of-fit test for standard AR(1) models in section \ref{examples}.
Further we reconsider the test for multiplicative structure by Dette, Pardo-Fernßndez \& Van Keilegom (2009) under the normality assumption. 

Finally, technical assumptions are listed in appendix \ref{section-tech}. 

\section{Main results}
\def\theequation{2.\arabic{equation}}
\setcounter{equation}{0}

\subsection{AR-ARCH model}\label{section-AR-ARCH}

Assume we have observed $X_0,\ldots,X_n$, where $(X_t)_{t\in\zet}$ is a real valued stationary $\alpha$-mixing stochastic process following the AR-ARCH model of order one, i,\,e.\
\begin{eqnarray}X_t=m(X_{t-1})+\sigma(X_{t-1})\e_t.\label{mod}\end{eqnarray}
Here the innovations $\e_t$, $t\in\zet$, are assumed independent and identically distributed with unknown distribution function $F$. Moreover, the innovations are centered with unit variance and $\e_t$ is independent of the past $X_s$, $s\leq t-1$, $\forall t$.

Our aim is to test the null hypothesis $H_0$ of standard normal innovations against a general alternative $H_1$. To this end we define kernel estimators for the conditional mean and conditional variance function as
\begin{eqnarray}\label{m-sig}
\hat{m}(x)&=&\frac{\sum_{i=1}^nK\left(\frac{x-X_{i-1}}{c_n}\right)X_i}{\sum_{i=1}^nK\left(\frac{x-X_{i-1}}{c_n}\right)}\ ,\quad \hat{\sigma}^2(x)\;=\;\frac{\sum_{i=1}^nK\left(\frac{x-X_{i-1}}{c_n}\right)(X_i-\hat m(x))^2}{\sum_{i=1}^nK\left(\frac{x-X_{i-1}}{c_n}\right)}\quad
\end{eqnarray}
where $K$ denotes a kernel function and $c_n$ a sequence of positive bandwidths. Technical assumptions are listed in the appendix. Now we estimate the innovations as residuals 
\[\he_t=\frac{X_t-\hat{m}(X_{t-1})}{\hat{\sigma}(X_{t-1})}\]
and consider a weighted empirical distribution function, i.\,e. 
\begin{eqnarray}\label{Fn}
\hat{F}_n(y)&=&\sum_{t=1}^nv_{n,t}I\{\hat{\e}_t\leq y\},
\end{eqnarray}
as estimator for the innovation distribution. Here we define
$$v_{n,t}=\frac{w_{n}(X_{t-1})}{\sum_{s=1}^nw_{n}(X_{s-1})}$$
while $w_n$ denotes some weight function fulfilling assumption (W) in the appendix.  
Selk \& Neumeyer (2012) showed (see (A.1), (A.3) and the arguments following in the proof of Th.\ 3.1 in that paper) that under the assumptions stated in the appendix, 
\begin{eqnarray}
\nn \hat F_n(y)&=&\sum_{t=1}^nv_{n,t}I\Big\{{\e}_t\leq \frac{(\hat m-m)(X_{t-1})}{\sigma(X_{t-1})}+y\frac{\hat\sigma(X_{t-1})}{\sigma(X_{t-1})}\Big\}\\
\label{Fn-ent}&=& \frac 1n\sum_{t=1}^nI\{\e_t\leq y\}+f(y)\sum_{t=1}^nv_{n,t}\Big(\frac{(\hat m-m)(X_{t-1})}{\sigma(X_{t-1})}+y\frac{(\hat\sigma^2-\sigma^2)(X_{t-1})}{2\sigma^2(X_{t-1})}\Big) \quad\\
&&{}+\op\nn
\end{eqnarray}
uniformly with respect to $y\in\er$, where $f$ denotes the innovation density. Further in the aforementioned paper it is shown that
\begin{eqnarray}\label{m-ent}
\sum_{t=1}^nv_{n,t}\frac{(\hat m-m)(X_{t-1})}{\sigma(X_{t-1})}&=& \frac{1}{n}\sum_{t=1}^n\e_t+\op\\
\sum_{t=1}^nv_{n,t}\frac{(\hat\sigma^2-\sigma^2)(X_{t-1})}{2\sigma^2(X_{t-1})}
&=& \frac{1}{2n}\sum_{t=1}^n (\e_t^2-1) +\op \label{sig-ent}
\end{eqnarray}
(see (A.5)--(A.7) in the cited paper). 
Thus,
\begin{eqnarray}\label{Num}
\hat F_n(y)&=& \frac 1n\sum_{t=1}^n \left( I\left\{\e_t\leq y\right\}+f(y)\e_t+\frac{yf(y)}{2} (\e_t^2-1 )\right)+\op
\end{eqnarray}
and the stochastic process 
\[\sqrt n \left(\hat F_n(y)-F(y)\right),\;y\in\er,\]
converges weakly to a centered Gaussian process $(\ka(y))_{y\in\er}$
with \rm
\beq
\text{Cov}(\ka(y),\ka(z))&=&F(y\wedge z)-F(y)F(z)\\
&&{}+f(y)\left(E[\e_1I\{\e_1\leq z\}]+yE[(\e_1^2-1)I\{\e_1\leq z\}]\right)\\
&&{}+f(z)\left(E[\e_1I\{\e_1\leq y\}]+zE[(\e_1^2-1)I\{\e_1\leq y\}]\right)\\
&&{}+f(y)f(z)\left(1+(y+z)E[\e_1^3]+yz(E[\e_1^4]-1)\right).
\eeq
Now let $\Phi$ and $\varphi$ denote distribution and density function of the standard normal distribution, respectively, and denote by $(\ge(u))_{u\in[0,1]}$ a centered Gaussian process with covariance structure
\beq
\text{Cov}\left(\ge(u),\ge(v)\right)
&=&u\wedge v-uv-\varphi(\Phi^{-1}(u))\varphi(\Phi^{-1}(v)).
\eeq
Then we have the following result for the CramÚr-von Mises test statistic.

\begin{theo}\label{theo1} Under model (\ref{mod}) and the assumptions stated in the appendix under the null hypothesis $H_0$ of Gaussian innovations the test statistic 
$$T_n=n\int_\er (\hat F_n(y)-\Phi(y))^2\varphi(y)\,dy$$
converges in distribution to $\ds T=\int_0^1 \ge^2(u)\,du$.
\end{theo}

{\bf Proof.} A calculation of the covariance of $\ka$ in the case $F=\Phi$ gives 
\beq
\text{Cov}(\ka(y),\ka(z))&=&\Phi(y\wedge z)-\Phi(y)\Phi(z)\\
&&{}-2\varphi(y)\varphi(z)-2yz\varphi(y)\varphi(z)+\varphi(y)\varphi(z)(1+2yz)\\
&=&\Phi(y\wedge z)-\Phi(y)\Phi(z)-\varphi(y)\varphi(z).
\eeq
because for $\e_1$ standard normally distributed one easily calculates $E[\e_1 I\{\e_1\leq y\}]=-\varphi(y)$, 
$E[(\e_1^2-1) I\{\e_1\leq y\}]=-y\varphi(y)$ and one has $E[\e_1^3]=0$, $E[\e_1^4]=3$.
From the continuous mapping theorem it follows that $T_n$ converges in distribution to
$$\int_\er {\ka(y)}^2\varphi(y)\, dy=\int_{0}^{1}\left(\ka(\Phi^{-1}(u)) \right)^2\, du$$
while $\ka\circ \Phi^{-1}$ has the same distribution as $\ge$. This finishes the proof. 
\hfill $\Box$

\begin{rem}\rm 
It follows from Stephens (1976) that $\ge$ is also the weak limit of some process $(Y(u))_{u\in[0,1]}$ such that  $\tilde T=\int_0^1Y^2(u)du$ is the limit of 
$$ \tilde T_n=n\int\left(\frac 1n\sjn I\left\{Z_j\leq \cdot\right\}-\Phi_{\hat{\mu},\tau^2}\right)^2\,d\Phi_{\hat{\mu},\tau^2},$$
where $Z_1,\ldots,Z_n$ are iid with known variance $\tau^2$ and unknown expectation $\mu$ and where $\Phi_{\hat{\mu},\tau^2}$ denotes the normal distribution function with expectation $\hat{\mu}=n^{-1}\sum_{j=1}^n Z_j$ and variance $\tau^2$.
That the limits of $T_n$ and $\tilde T_n$ coincide in distribution might be suprising because in our AR-ARCH model the variance is unknown and has to be estimated. However, as can be seen from the proof, in the asymptotic covariance of $\sqrt{n}(\hat F_n-\Phi)$ under $H_0$ exactly those terms cancel that arise from the estimation of the variance function $\hat \sigma^2$ (cf.\ (\ref{sig-ent})). 
Quantiles of $\tilde T$ and thus critical values for $T$ are tabled in Stephens (1976) and restated in Table \ref{tab1} for convenience. We obtain an asymptotically distribution-free test by rejecting $H_0$ for asymptotic level $\alpha$ whenever $T_n$ is larger than the critical value $c_\alpha$. Consistency can be deduced from uniform convergence of $\hat F_n$ to $F$ in probability, which follows from (\ref{Num}). 
$\blacksquare$
\end{rem}

\vspace{0.5cm}

\begin{table}[h]\begin{center}
\begin{tabular}{| l | l | l | l | l | l | }
\hline
nominal level  $\alpha$ &$0.15$&$0.1$&$0.05$&$0.025$&$0.01$\\
\hline
critical value $c_\alpha$ &$ 0.118$&$ 0.135 $&$ 0.165 $&$ 0.196 $&$ 0.237 $\\
\hline
\end{tabular}
\caption{ \label{tab1} \sl Asymptotic critical values for the CramÚr-von Mises test for normality of the innovations in the AR-ARCH model.}
\end{center}
\end{table}

\vspace{0.5cm}

\begin{rem}\label{rem-gen}\rm 
The results for the nonparametric AR-ARCH model (\ref{mod}) can be extended to models of the form $X_t=m(Z_t)+\sigma(Z_t)\e_t$ where $(X_t,Z_t)$ is a stationary time series and $Z_t$ may include covariates as well as a finite number of past values $X_{t-1},\dots,X_{t-p}$ while $E[\e_t\mid \mathcal{F}_{t-1}]=0$, $\Var(\e_t\mid \mathcal{F}_{t-1})=1$. Here $\mathcal{F}_{t-1}$ denotes the sigma-field generated by $Z_t,(X_{t-1},Z_{t-1}),(X_{t-2},Z_{t-2}),\dots$.
We conjecture that applying local polynomial estimators for $m$ and $\sigma^2$ and assuming enough smoothness of those functions  the expansion (\ref{Num}) stays valid.  A thorough treatment is beyond the scope of the paper. Note also that asymptotic properties of estimators for the innovation distribution in such nonparametric AR($p$)/regression models have not yet been treated in the literature. 
However, in the case of independent observations Neumeyer \& Van Keilegom (2010) showed validity of an expansion like (\ref{Num}) for the empirical distribution of residuals in multiple nonparametric regression models (they obtain the same expansion as in the case of one-dimensional covariates, see Akritas \& Van Keilegom (2001)).  Thus we believe that Theorem \ref{theo1} stays valid in the more general model under suitable regularity conditions and the same test for Gaussianity of the innovation distribution can be applied. $\blacksquare$
\end{rem}

\subsection{AR model}\label{section-AR}

In this section we consider a nonparametric AR model of order one, i.\,e.\
\begin{eqnarray}X_t=m(X_{t-1})+\eta_t, \label{mod-AR}\end{eqnarray}
where the innovations $\eta_t$, $t\in\zet$, are iid and centered and $\eta_t$ is independent of the past $X_s$, $s\leq t-1$. Thus we have model (\ref{mod}) with $\eta_t=\sigma\e_t$ for the unknown (constant) variance $\sigma^2=\Var(\eta_t)$. Our aim is to test the null hypothesis of normal innovations, i.\,e.\
$$H_0:\exists\sigma^2>0 \mbox{ s.\,t.\ } \eta_t\sim N(0,\sigma^2).$$
The constant variance is estimated by
\begin{eqnarray*}
\hat\sigma^2&=& \sum_{t=1}^n v_{n,t}(X_t-\hat m(X_{t-1}))^2,
\end{eqnarray*}
where $\hat m$ is the kernel estimator defined in (\ref{m-sig}). 
In this case we define the residuals as
\[\he_t=\frac{X_t-\hat{m}(X_{t-1})}{\hat{\sigma}}\]
and consider $\hat F_n$ as defined in (\ref{Fn}) as estimator for the distribution of the standardized innovations $\e_t$. Let again $F$ and $f$ denote distribution and density function of $\e_t$, respectively. Then $H_0$ is equivalent to $F=\Phi$ and we have the following results.

\begin{lemma}\label{lem1} Under the assumptions stated in the appendix we have the expansion
\beq
\hat F_n(y)&=&\frac 1n\sum_{t=1}^n \left( I\left\{\e_t\leq y\right\}+f(y)\e_t+\frac{yf(y)}{2} (\e_t^2-1 )\right)+\op
\eeq
uniformly with respect to $y\in\er$. 
\end{lemma}

{\bf Proof.} First we consider the variance estimator, for which one obtains
\begin{eqnarray*}
\hat\sigma^2-\sigma^2&=& \frac{1}{n}\sum_{t=1}^n v_{n,t}(\eta_t-(\hat m-m)(X_{t-1}))^2-\sigma^2\\
&=& \sigma^2\frac{1}{n}\sum_{t=1}^nv_{n,t}(\e_t^2-1)-2\sigma\sum_{t=1}^nv_{n,t}\e_t(\hat m-m)(X_{t-1})+\sum_{t=1}^nv_{n,t}(\hat m-m)^2(X_{t-1})\\
&=& \sigma^2\sum_{t=1}^nv_{n,t}(\e_t^2-1) +\op
\end{eqnarray*}
by Lemmata B.2 and B.3 in Selk \& Neumeyer (2012).
Further 
$$\frac 1n\sum_{t=1}^n (w_n(X_{t-1})-1)(\e_t^2-1)=\op$$
and
$$\frac 1n\sum_{t=1}^n w_n(X_{t-1})=1+\op$$
(see also (A.4) in Selk \& Neumeyer (2012)) and thus
\begin{eqnarray}
\label{sig-ent2}
\hat\sigma^2-\sigma^2&=& \sigma^2\frac 1n\sum_{t=1}^n(\e_t^2-1)+\op.
\end{eqnarray}

Now note that
\begin{eqnarray*}
\hat{F}_n(y)&=&\sum_{t=1}^nv_{n,t}I\Big\{{\e}_t\leq \frac{(\hat m-m)(X_{t-1})}{\sigma}+y\frac{\hat\sigma}{\sigma}\Big\}
\end{eqnarray*}
and with arguments analogous to the proof of Theorem 3.1 in Selk \& Neumeyer (2012) which leads to (\ref{Fn-ent}) (see also the proof of Lemma 1 in Dette, Pardo-Fernßndez \& Van Keilegom (2009) or the proof of Theorem 3.1 in M³ller, Schick \& Wefelmeyer (2009)) one obtains that
\begin{eqnarray*}
\hat{F}_n(y)&=&\frac 1n\sum_{t=1}^nI\{\e_t\leq y\}+f(y)\sum_{t=1}^nv_{n,t}\Big(\frac{(\hat m-m)(X_{t-1})}{\sigma}+y\frac{\hat\sigma^2-\sigma^2}{2\sigma^2}\Big) \\
&&{}+\op
\end{eqnarray*}
uniformly with respect to $y\in\er$. The assertion follows from (\ref{m-ent}) and  (\ref{sig-ent2}). 
\hfill $\Box$

\begin{cor} Under model (\ref{mod-AR}) and the assumptions stated in the appendix under the null hypothesis $H_0$ of normal innovations
$$T_n=n\int_\er (\hat F_n(y)-\Phi(y))^2\varphi(y)\,dy$$
converges in distribution to $T$ defined in Theorem \ref{theo1}.
\end{cor}

{\bf Proof.} The result immediately follows from Lemma \ref{lem1} and Theorem \ref{theo1}.
\hfill $\Box$

\medskip

Thus using the critical values from Table \ref{tab1} we obtain a completely nonparametric consistent distribution-free asymptotic level $\alpha$ test for Gaussianity of the innovations in AR models (see also Remark \ref{rem-gen} which holds here analogously).

\subsection{ARCH model}\label{section-ARCH}

In this section we consider a nonparametric ARCH model of order one, i.\,e.\
\begin{eqnarray}\label{mod-ARCH}
X_t=\sigma(X_{t-1})\e_t, 
\end{eqnarray}
where the innovations $\e_t$, $t\in\zet$, are iid and centered with unit variance and $\e_t$ is independent of the past $X_s$, $s\leq t-1$. Thus we have model (\ref{mod}) with conditional mean $m\equiv 0$. Our aim is to test the null hypothesis $H_0$ of normal innovations. To this end let $\sigma^2$ be estimated by the kernel estimator \[\hat{\sigma}^2(x)=\;\frac{\sum_{i=1}^nK\left(\frac{x-X_{i-1}}{c_n}\right)X_i^2}{\sum_{i=1}^nK\left(\frac{x-X_{i-1}}{c_n}\right)}\] and the residuals be defined as
\[\he_t=\frac{X_t}{\hat{\sigma}(X_{t-1})},\]
whereas $F_n$ is as in (\ref{Fn}). Let again $\Phi$ denote the standard normal distribution function and let $\be$ denote a standard Brownian bridge on $[0,1]$. Then we obtain the following result. 

\begin{theo}\label{theo2} Under the assumptions stated in the appendix under the null hypothesis of Gaussian innovations in the ARCH model (\ref{mod-ARCH})  the test statistic 
$$T_n=n\int_\er (\hat F_n(y)-\Phi(y))^2\varphi(y)\,dy$$
converges in distribution to $\ds T=\int_0^1 \be^2(u)\,du$.
\end{theo}

{\bf Proof.} In the expansion (\ref{Fn-ent}) the estimation of $m$ is not present while (\ref{sig-ent}) stays valid for the new estimator $\hat{\sigma}$. Thus it follows that 
\begin{eqnarray*}
\hat F_n(y)&=& \frac 1n\sum_{t=1}^n \left( I\left\{\e_t\leq y\right\}+\frac{yf(y)}{2} (\e_t^2-1 )\right)+\op
\end{eqnarray*}
and the stochastic process 
\[\sqrt n \left(\hat F_n(y)-F(y)\right),\;y\in\er,\]
converges weakly to a centered Gaussian process $(\ka(y))_{y\in\er}$
with 
\beq
\text{Cov}(\ka(y),\ka(z))&=&F(y\wedge z)-F(y)F(z)\\
&&{}+f(y)yE[(\e_1^2-1)I\{\e_1\leq z\}]+f(z)zE[(\e_1^2-1)I\{\e_1\leq y\}]\\
&&{}+f(y)f(z)yz(E[\e_1^4]-1).
\eeq
As the innovations are standard normally distributed we have $E[(\e_1^2-1)I\{\e_1\leq z\}]=-z\varphi(z)$, $f=\varphi$ and $E[\e_1^4]=3$. Hence, 
\beq
\text{Cov}(\ka(y),\ka(z))&=&\Phi(y\wedge z)-\Phi(y)\Phi(z).
\eeq
The assertion follows by the continuous mapping theorem noting that $\ka\circ\Phi^{-1}$ is a standard Brownian bridge. 
\hfill $\Box$

\medskip

For convenient reference we state the critical values for the test in Table \ref{tab2}. Here $c_\alpha$ is the ($1-\alpha$)-quantile of $T=\int_0^1 \be^2(u)\,du$, see Shorack \& Wellner (1986), p.\,147.

\vspace{0.5cm}

\begin{table}[h]\begin{center}
\begin{tabular}{| l | l | l | l | l | l | }
\hline
nominal level  $\alpha$ &$0.15$&$0.1$&$0.05$&$0.02$&$0.01$\\
\hline
critical value $c_\alpha$ & $0.284$& $0.347$&$0.461$&$0.6198 $&$0.743$\\
\hline
\end{tabular}
\caption{ \label{tab2} \sl Asymptotic critical values for the CramÚr-von Mises test for normality of the innovations in the ARCH model.}
\end{center}
\end{table}


\section{Simulations}\label{section-simus}
\def\theequation{3.\arabic{equation}}
\setcounter{equation}{0}

\begin{figure}[h!]\begin{center}
\begin{minipage}[t]{0.47\textwidth}
\includegraphics[width=\textwidth]{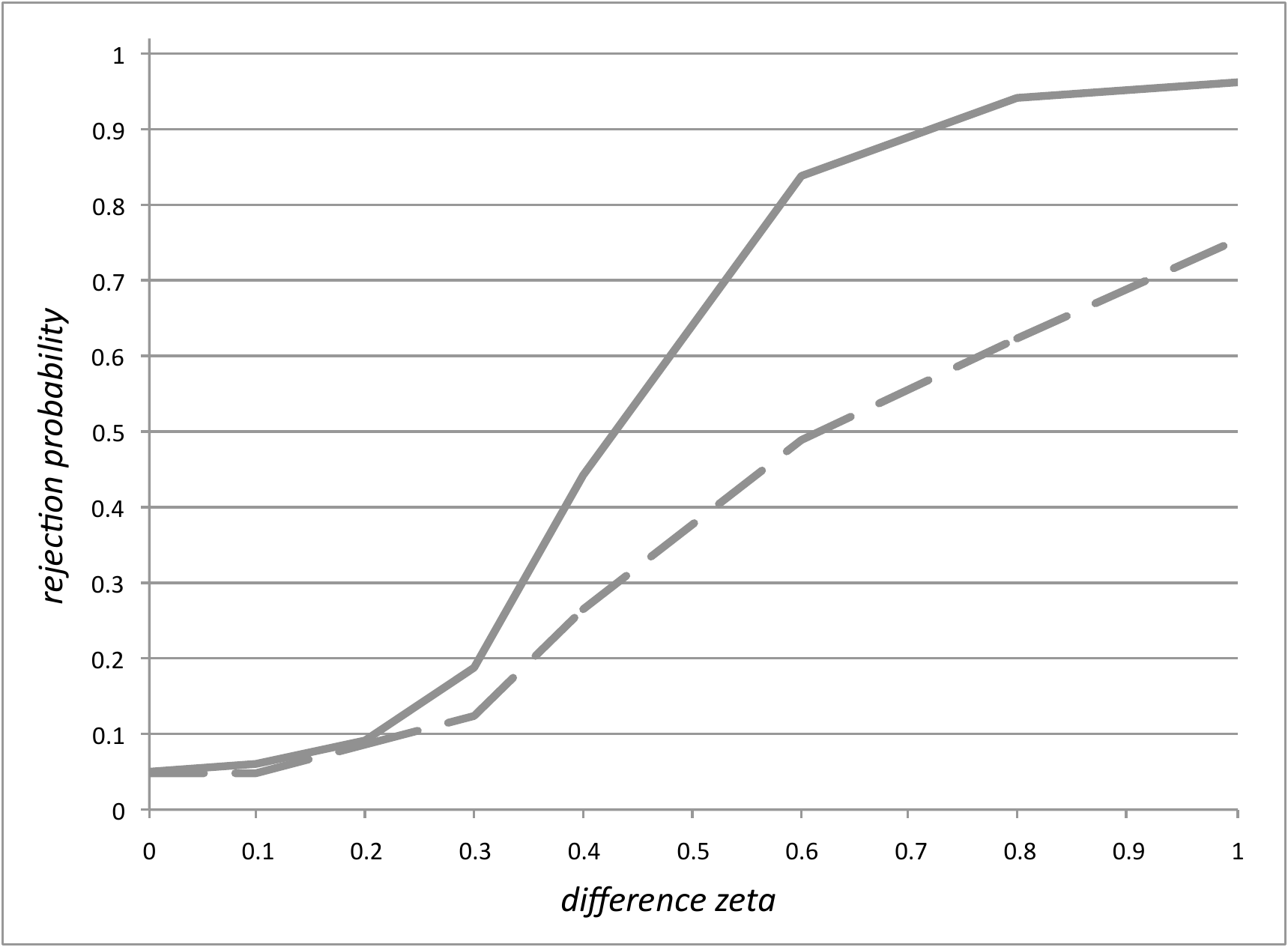}
\end{minipage}
\quad
\begin{minipage}[t]{0.47\textwidth}
\includegraphics[width=\textwidth]{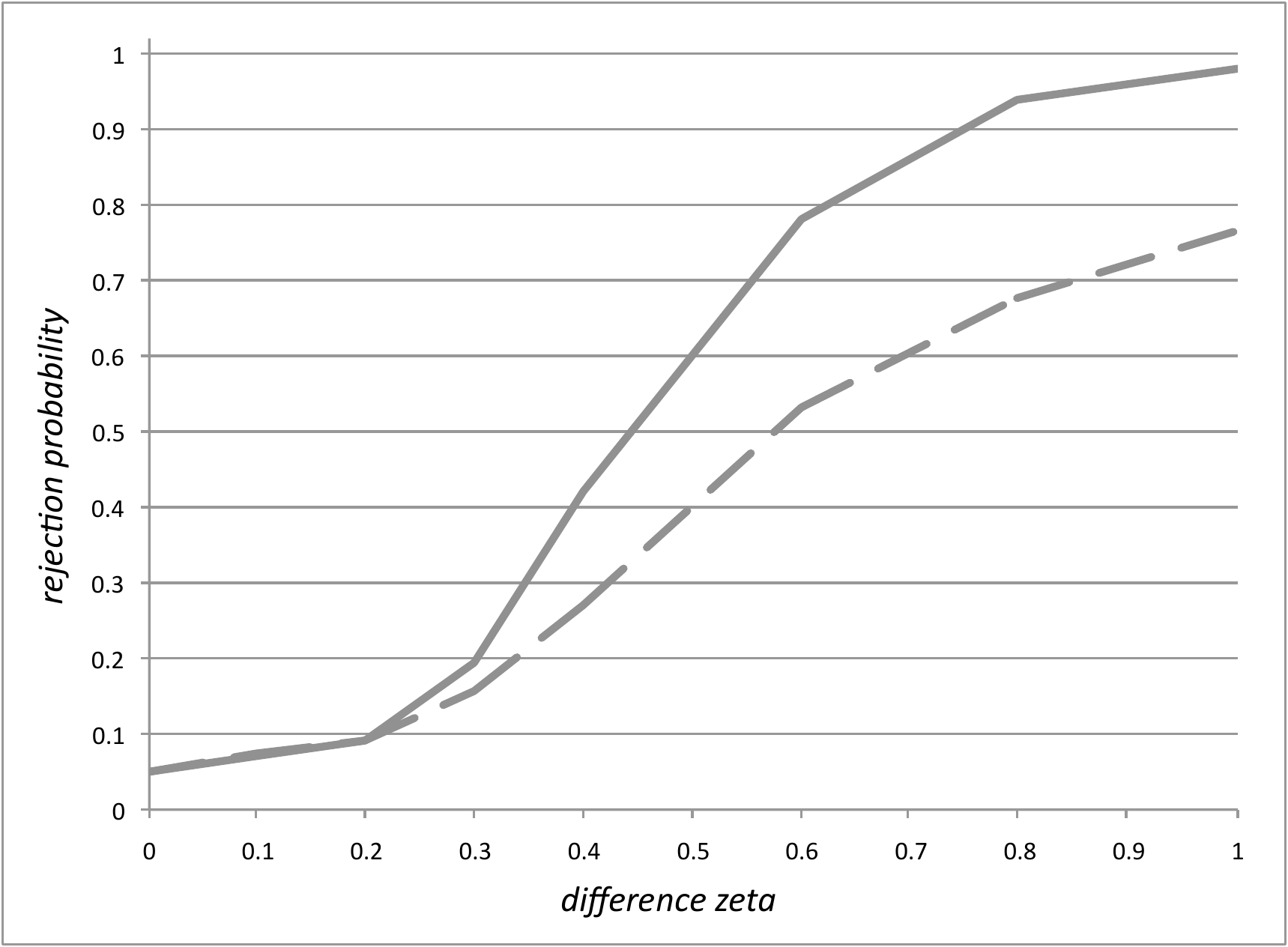}
\end{minipage}
\caption{\sl Rejection probabilities obtained from AR(1) models with skew-normally distributed innovations for $n=100$ (dashed curve) and $n=200$ (solid curve). On the left panel the results for the AR-ARCH model are shown and on the right the ones for the AR model.}\label{ar1snb}
\end{center}\end{figure}

To examine the performance of the test on small samples we consider AR(1) models and ARCH(1) models, for which we compare the results under the assumption of an AR-ARCH model like (\ref{mod}) and under the assumption of an AR model like (\ref{mod-AR}) (respectively ARCH like (\ref{mod-ARCH})).\\
For the AR(1) case we consider the models
\[X_t=0.5\cdot X_{t-1}+\e_t,\qquad \e_1,\ldots,\e_n\sim\tilde{F}_{\zeta},\]
where $\tilde F_{\zeta}$ denotes the skew-normal distribution with location parameter $$-\sqrt{\frac{2\pi\left(\left(5\zeta\right)^2+\left(5\zeta\right)^4\right)}{\pi^2+\left(2\pi^2-2\pi\right)\cdot\left(5\zeta\right)^2+\left(\pi^2-2\pi\right)\cdot\left(5\zeta\right)^4}},$$ scale parameter
$(\pi(1+(5\zeta)^2)^{1/2}/(\pi+(\pi-2)(5\zeta)^2)^{1/2}$ and shape parameter $5\zeta$ for different values of $\zeta$. For $\zeta=0$ this is the standard normal distribution.
The rejection probabilities for 500 repetitions and level 5\% are displayed in Table \ref{ar1sn} and Figure \ref{ar1snb} for the AR-ARCH model (\ref{mod}) and the AR model (\ref{mod-AR})  respectively. It can be seen that the level is approximated well and the power increases for increasing parameter $\zeta$ as well as for increasing sample size $n$.

\begin{table}[h!]\begin{center}
\begin{tabular}{| l || c | c | c | c | c | c | c | c |}
\hline
 $\quad\%$&$\ \zeta=0\ $&$\zeta=0.1$&$\zeta=0.2$&$\zeta=0.3$&$\zeta=0.4$&$\zeta=0.6$&$\zeta=0.8$&$\zeta=1$\\
\hline\hline
AR-ARCH &&&&&&&&\\
\hline
$n=100$&$4.8$&$5.4$&$8.6$&$12.2$&$26.2$&$48.8$&$62$&$75.4$\\
\hline
$n=200$&$5$&$6$&$8.8$&$18.6$&$44$&$83.6$&$94$&$96.2$\\
\hline\hline
AR &&&&&&&&\\
\hline
$n=100$&$5$&$7.4$&$9$&$15.6$&$27$&$53.2$&$67.4$&$76.6$\\
\hline
$n=200$&$5$&$6.8$&$8.8$&$19$&$41.8$&$77.8$&$93.6$&$98$\\
\hline
\end{tabular}
\caption{\sl Rejection probabilities obtained from AR(1) models with skew-normally distributed innovations}\label{ar1sn}
\end{center}\end{table}

\begin{figure}[h!]\begin{center}
\begin{minipage}[t]{0.47\textwidth}
\includegraphics[width=\textwidth]{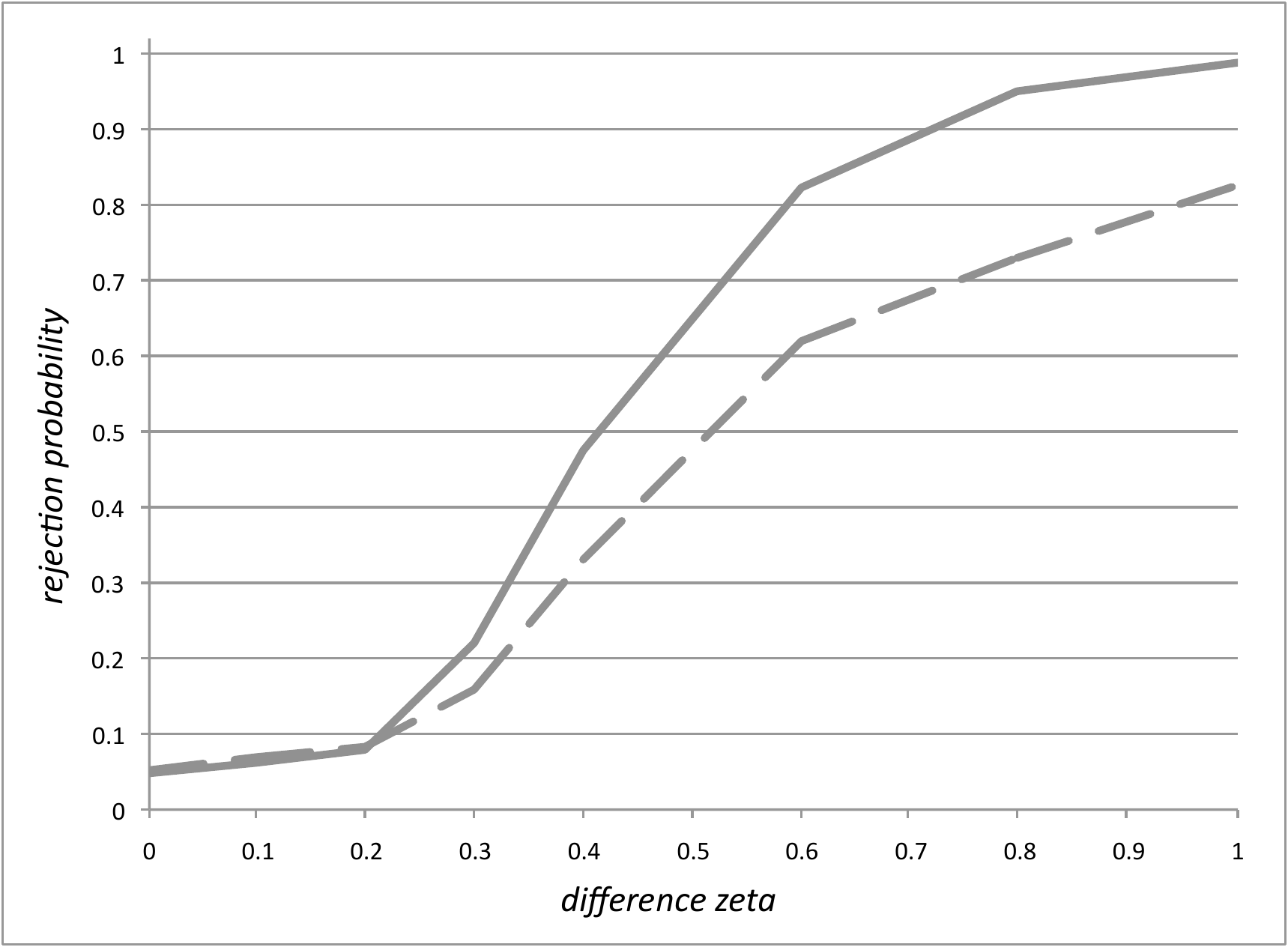}
\end{minipage}
\quad
\begin{minipage}[t]{0.47\textwidth}
\includegraphics[width=\textwidth]{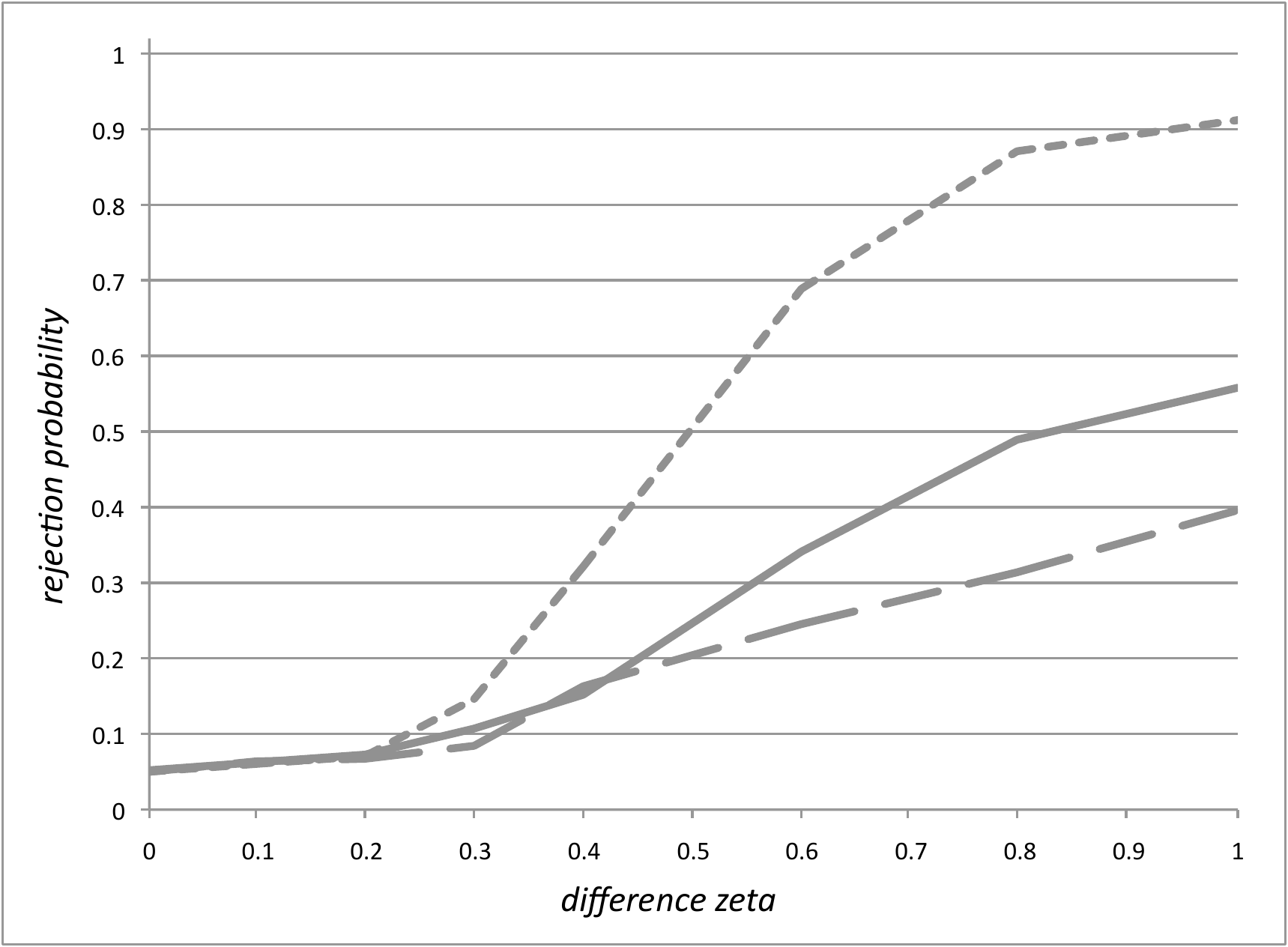}
\end{minipage}
\caption{\sl Rejection probabilities obtained from ARCH(1) models with skew-normally distributed innovations for $n=100$ (dashed curve), $n=200$ (solid curve) and $n=500$ (dotted curve). On the left panel the results for the AR-ARCH model are shown and on the right panel the ones for the ARCH model.}\label{arch1snb}
\end{center}\end{figure}

We also examine ARCH(1) models with the same innovation distribution,
\[X_t=\sqrt{0.75+0.25X_{t-1}^2}\cdot\e_t,\qquad \e_1,\ldots,\e_n\sim\tilde{F}_{\zeta}\]
for different values of $\zeta$. 
The rejection probabilities for 500 repetitions and level 5\% are shown in Table \ref{arch1sn} and Figure \ref{arch1snb} for the AR-ARCH model (\ref{mod}) and the ARCH model (\ref{mod-ARCH})  respectively. The asymptotic level is approximated well and the power increases with increasing $\zeta$ as well as with increasing $n$. For the ARCH model the increase with $\zeta$ for small $n$ is not as pronounced as for the models considered before, therefore we additionally examined this model with sample size $n=500$ for which an increase comparable to those before can be observed.

\begin{table}[h!]\begin{center}
\begin{tabular}{| l || c | c | c | c | c | c | c | c |}
\hline
 $\quad\%$&$\ \zeta=0\ $&$\zeta=0.1$&$\zeta=0.2$&$\zeta=0.3$&$\zeta=0.4$&$\zeta=0.6$&$\zeta=0.8$&$\zeta=1$\\
\hline\hline
AR-ARCH &&&&&&&&\\
\hline
$n=100$&$5.2$&$6.6$&$8.2$&$15.8$&$33$&$61.8$&$72.8$&$82.6$\\
\hline
$n=200$&$4.8$&$6.2$&$7.8$&$22$&$47.4$&$82.2$&$94.8$&$98.8$\\
\hline\hline
ARCH &&&&&&&&\\
\hline
$n=100$&$5$&$6.2$&$6.4$&$8$&$16.2$&$24.4$&$31.2$&$39.6$\\
\hline
$n=200$&$5.2$&$6$&$7.2$&$10.4$&$15$&$34$&$48.8$&$55.8$\\
\hline
$n=500$&$5$&$5.6$&$6.8$&$14.2$&$31.8$&$68.6$&$87$&$91.2$\\
\hline
\end{tabular}
\caption{\sl Rejection probabilities obtained from ARCH(1) models with skew-normally distributed innovations}\label{arch1sn}
\end{center}\end{table}

\begin{figure}[h!]\begin{center}
\begin{minipage}[t]{0.47\textwidth}
\includegraphics[width=\textwidth]{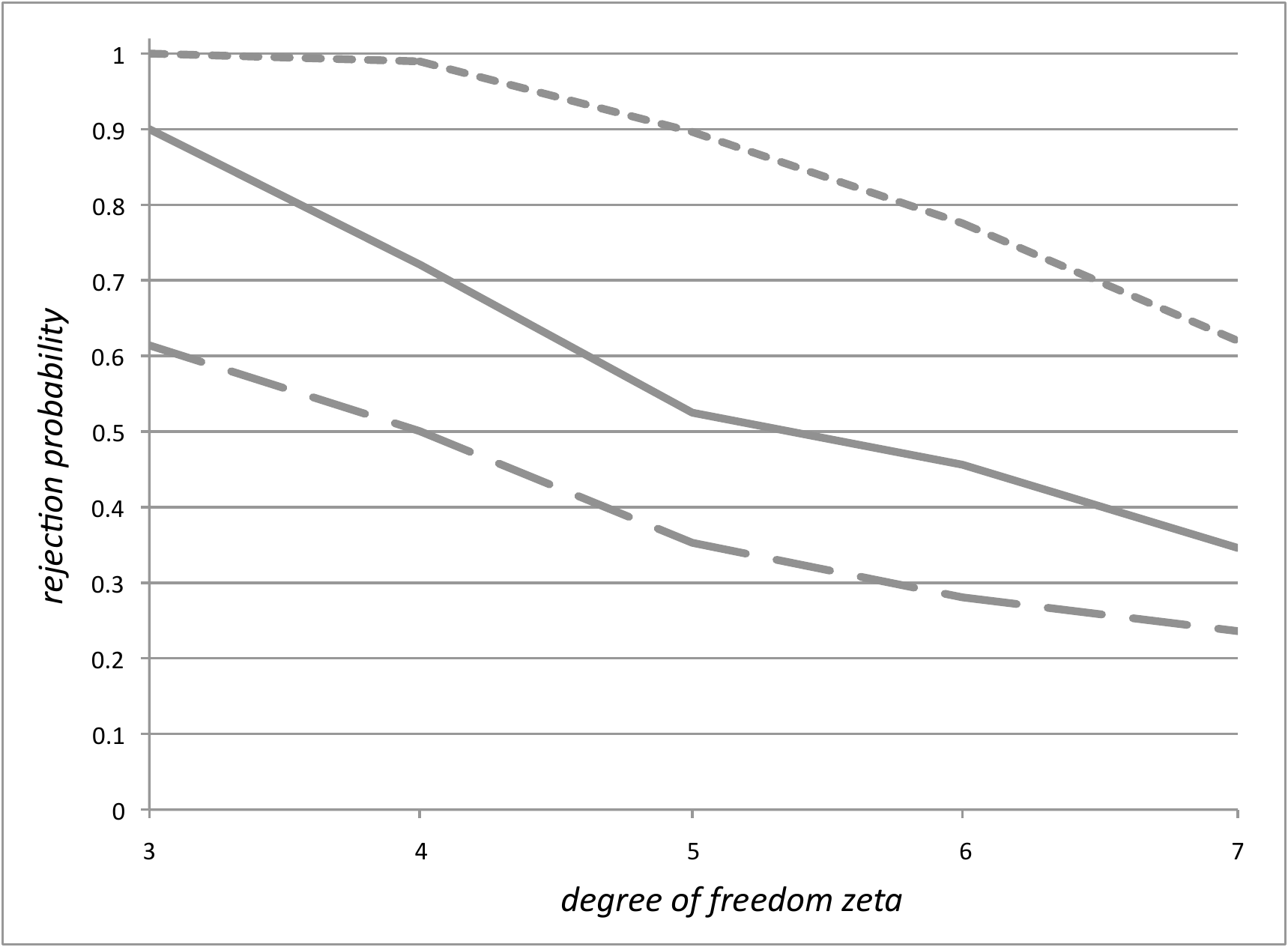}
\end{minipage}
\quad
\begin{minipage}[t]{0.47\textwidth}
\includegraphics[width=\textwidth]{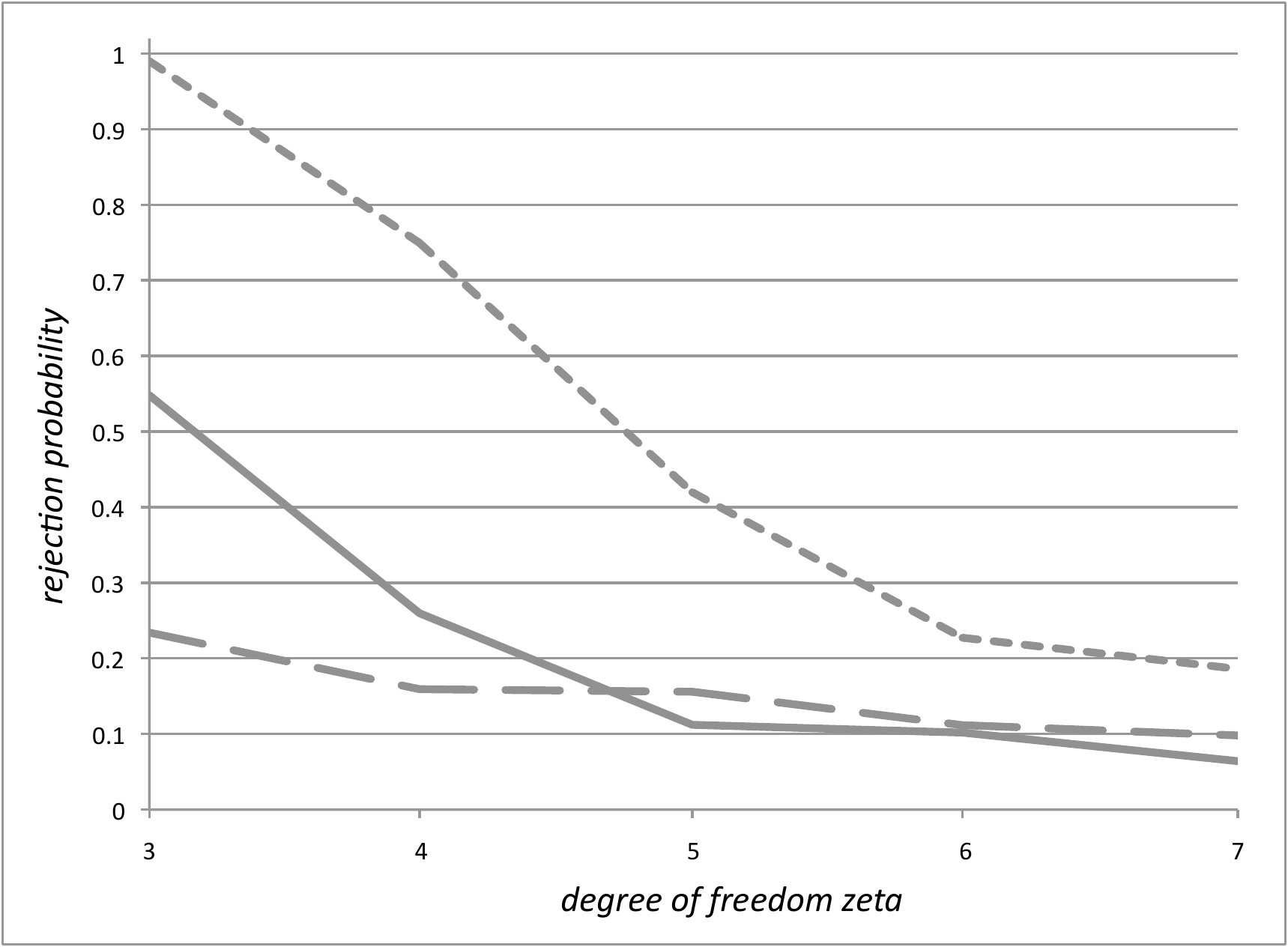}
\end{minipage}
\caption{\sl Rejection probabilities obtained from AR(1) models with $\mathcal{S}t(\zeta)$ distributed innovations for $n=100$ (dashed curve), $n=200$ (solid curve) and $n=500$ (dotted curve). On the left panel the results for the AR-ARCH model are shown and on the right panel the ones for the AR model.}\label{ar1tb}
\end{center}\end{figure}

\begin{figure}[h!]\begin{center}
\begin{minipage}[t]{0.47\textwidth}
\includegraphics[width=\textwidth]{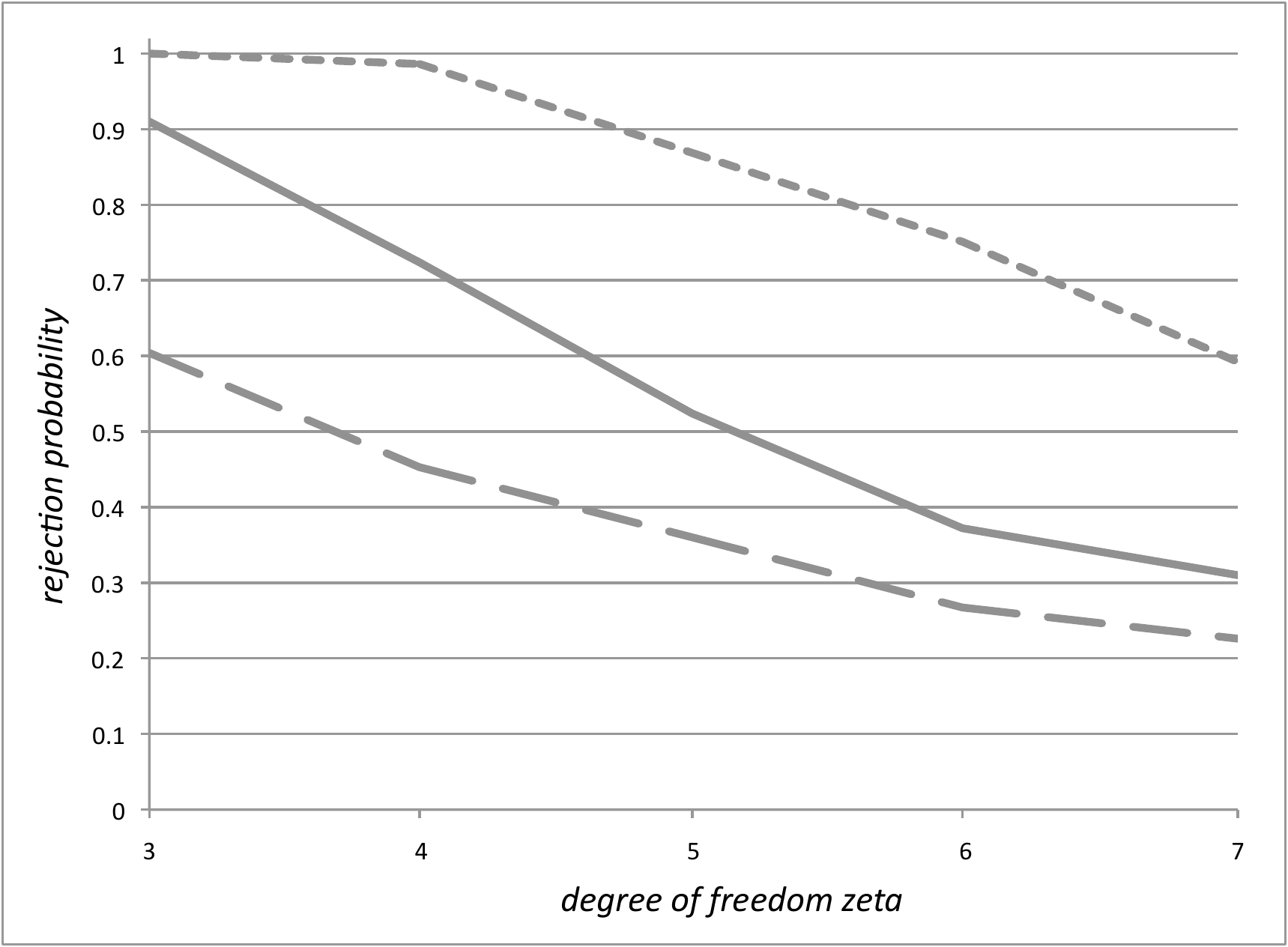}
\end{minipage}
\quad
\begin{minipage}[t]{0.47\textwidth}
\includegraphics[width=\textwidth]{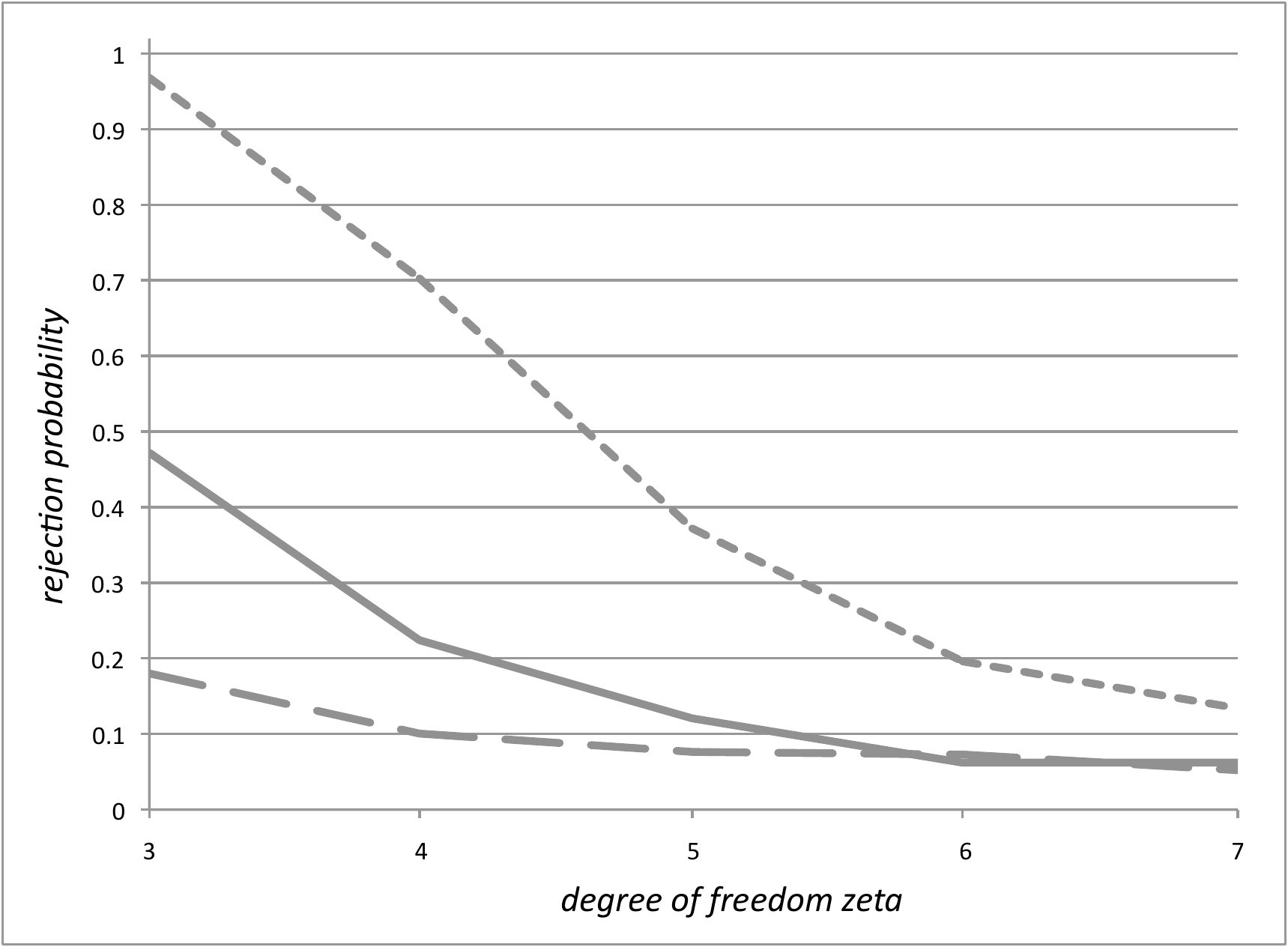}
\end{minipage}
\caption{\sl Rejection probabilities obtained from ARCH(1) models with $\mathcal{S}t(\zeta)$ distributed innovations for $n=100$ (dashed curve), $n=200$ (solid curve) and $n=500$ (dotted curve). On the left panel the results for the AR-ARCH model are shown and on the right panel the ones for the ARCH model.}\label{arch1tb}
\end{center}\end{figure}

To further study the power of the testing procedure, we examine the same models with Student-t distributed innovations with different degrees of freedom.
Due to the fact that Var$(\e_t)$ has to be one for all $t$, the Student-t distribution was standardized.
The models are AR(1)
\[X_t=0.5\cdot X_{t-1}+\e_t,\qquad \e_1,\ldots,\e_n\sim\mathcal{S}t(\zeta)\]
and ARCH(1)
\[X_t=\sqrt{0.75+0.25X_{t-1}^2}\cdot\e_t,\qquad \e_1,\ldots,\e_n\sim\mathcal{S}t(\zeta)\]
for different values of $\zeta$.
The rejection probabilities for 500 repetitions and level 5\% are displayed in Table \ref{t}, Figure \ref{ar1tb} (AR(1) model) and Figure \ref{arch1tb} (ARCH(1) model). As before we compare the results under the assumption of an AR-ARCH model like (\ref{mod}) and under the assumption of an AR model like (\ref{mod-AR}) (respectively ARCH like (\ref{mod-ARCH})). It can be seen that the power is good and increases for increasing sample size $n$ while it decreases for increasing parameter $\zeta$, because the Student-t distribution converges to the standard normal distribution for increasing degree of freedom.

\begin{table}[h!]\begin{center}
\begin{tabular}{| l || c | c | c | c | c || c | c | c | c | c |}
\hline
 $\quad\%$&$ \zeta=3 $&$\zeta=4$&$\zeta=5$&$\zeta=6$&$\zeta=7$&$\zeta=3 $&$\zeta=4$&$\zeta=5$&$\zeta=6$&$\zeta=7$\\
\hline\hline
\small AR-ARCH &&&&&&&&&&\\
\hline
$n=100$&$61.4$&$49.8$&$35.2$&$27.8$&$23.6$&$60.4$&$45.2$&$35.8$&$26.4$&$22.6$\\
\hline
$n=200$&$90$&$71.8$&$52.2$&$45,4$&$34.6$&$91$&$72.2$&$52.4$&$37.2$&$31$\\
\hline
$n=500$&$100$&$98.8$&$89.6$&$77.6$&$62$&$100$&$98.4$&$86.6$&$75$&$59.2$\\
\hline\hline
\small AR/ARCH &&&&&&&&&&\\
\hline
$n=100$&$23.4$&$15.8$&$15.4$&$11$&$9.8$&$18$&$9.8$&$7.6$&$7.2$&$5.2$\\
\hline
$n=200$&$54.8$&$25.6$&$10.8$&$9.8$&$6.4$&$47.2$&$22.4$&$12$&$6.6$&$6.2$\\
\hline
$n=500$&$99$&$74.6$&$41.8$&$22.4$&$18.6$&$96.8$&$70.2$&$37$&$19.6$&$13.4$\\
\hline
\end{tabular}
\caption{\sl Rejection probabilities obtained from AR(1) models (left) and ARCH(1) models (right) with $\mathcal{S}t(\zeta)$ distributed innovations}\label{t}
\end{center}\end{table}

{\bf Simulation setting.}
For each simulation $10\cdot n$ observations $X_t$ were generated. For the test the last $n$ observations were used. This was done to ensure that the process is in balance.\\
The empirical processes were built with weight function $w_{n}=I_{[-\log(n), \log(n)]}$.
The Nadaraya-Watson estimators $\hat{m}$ and $\hat{\sigma}$ were calculated with Gaussian kernel. This is not compatible with all assumptions, e.\,g.\ the support of the kernel is not compact. However this has negligible effect on the simulations because the Gaussian kernel decreases exponentially fast at the tails. The bandwidth was chosen according to the assumptions by a rule of thumb as $\hat\sigma^2n^{-\frac 2{6+\sqrt{3}}}$ with $\hat\sigma^2=\sum_{t=1}^nv_{n,t}(X_t-\hat m(X_{t-1}))^2$.


\section{Further examples}\label{examples}
\def\theequation{4.\arabic{equation}}
\setcounter{equation}{0}

\subsection{Testing for linear AR(1)}\label{section-AR(1)}

As was mentioned in the introduction other distribution-free specification tests for the AR-ARCH model can be derived once the Gaussianity of the errors has been established. 
In this section we will study in detail a lack-of-fit test for the linear AR(1) model. For the method compare Van Keilegom, Gonzßlez Manteiga, \& Sßnchez Sellero (2008) in a nonparametric regression model with independent observations. Similarly one can derive tests, e.\,g., for parametric ARCH models. 

For simplicity here we assume that $E[X_0]=0$ as this is a typical assumption in AR models. We consider the model 
\begin{eqnarray}X_t=m(X_{t-1})+\sigma\e_t,\label{mod-ex}\end{eqnarray}
where the innovations $\e_t$, $t\in\zet$, are iid standard normal and $\sigma$ is an unknown positive constant. Further,  $\e_t$ is independent of the past $X_s$, $s\leq t-1$. Our aim is to test the null hypothesis
$$\tilde H_0: \exists \vartheta\in (-1,1) \mbox{ s.\,t.\ } m(x)=\vartheta x.$$
See Hong-zhi \& Bing (1991) or Koul \& Stute (1999) for other procedures to test for $\tilde H_0$. 

Let $\hat\vartheta$ denote any (under $H_0$) $\sqrt{n}$-consistent estimator for $\vartheta$, e.\,g.\ the Yule-Walker or ordinary least squares estimator under suitable regularity assumptions. 
 Now define residuals under the null as
\[\he_{0,t}=\frac{X_t-\hat\vartheta X_{t-1}}{\hat{\sigma}}\]
with $\hat\sigma^2= n^{-1}\sum_{t=1}^n (X_t-\hat\vartheta X_{t-1})^2$,
 and
\begin{eqnarray*}
\hat{F}_{0,n}(y)&=&\sum_{t=1}^nI\{\hat{\e}_{0,t}\leq y\}
\end{eqnarray*}
 Let further $\hat F_n$ be defined as in section \ref{section-AR}.
Then we have the following result.

\begin{theo} Under the assumptions stated in the appendix for model (\ref{mod-ex}) with Gaussian innovations we have under $\tilde H_0$ that 
$$T_n=2\sqrt{\pi}n\int_\er (\hat F_n(y)-\hat F_{0,n}(y))^2\,dy$$
converges in distribution to a $\chi^2_1$-distributed random variable.
\end{theo}

{\bf Proof.} Let $\vartheta_0$ denote the `true' parameter under $H_0$. Similar to the proof of Lemma \ref{lem1} we have for standard Gaussian $\e_t$ that
\begin{eqnarray*}
\hat{F}_{0,n}(y)&=&\sum_{t=1}^nI\Big\{{\e}_t\leq \frac{(\hat\vartheta -\vartheta_0)(X_{t-1})}{\sigma}+y\frac{\hat\sigma}{\sigma}\Big\}\\
&=&\frac 1n\sum_{t=1}^nI\{\e_t\leq y\}+\varphi(y)\sum_{t=1}^n\Big(\frac{(\hat\vartheta-\vartheta_0)X_{t-1}}{\sigma}+y\frac{\hat\sigma^2-\sigma^2}{2\sigma^2}\Big) +\op
\end{eqnarray*}
uniformly with respect to $y\in\er$. Note that for the $\hat\sigma^2$ defined here the last equality in (\ref{sig-ent}) also holds. Further,
\begin{eqnarray*}
\frac 1n\sum_{t=1}^n X_{t-1} &=& E[X_0]+o_P(1)\;=\; o_P(1)
\end{eqnarray*}
because $(X_t)_{t\in\zet}$ is ergodic due to its mixing property (see e.g. Doukhan (1994)). Thus with the $\sqrt n$-consistency of  $\hat\vartheta$ we obtain
\begin{eqnarray*}
\hat{F}_{0,n}(y)&=&\frac 1n\sum_{t=1}^n\Big(I\{\e_t\leq y\}+\frac{y\varphi(y)}{2}(\e_t^2-1)\Big)+\op
\end{eqnarray*}
uniformly with respect to $y\in\er$. Now from Lemma \ref{lem1} we have
$$\sqrt{n}(\hat F_n(y)-\hat F_{0,n}(y))=\varphi(y)\frac{1}{\sqrt{n}}\sum_{t=1}^n \e_t+o_P(1)$$
which converges in distribution to $\varphi(y)Z$ with a standard normally distributed $Z$. Thus
$T_n$ converges in distribution to 
$$2\sqrt{\pi}\int_\er (\varphi(y))^2\,dy \,Z^2=Z^2, $$
which is $\chi^2_1$-distributed. 
\hfill $\Box$

\medskip

An asymptotically distribution-free level-$\alpha$ test is obtained by rejecting the null hypothesis $\tilde H_0$ of a standard AR(1)-model whenever $T_n$ is larger than the $(1-\alpha)$-quantile of the $\chi^2_1$-distribution.

\subsection{Testing for multiplicative structure}\label{section-mult}

Under the assumption of Gaussian innovations the test for multiplicative models
by Dette, Pardo-Fernßndez \& Van Keilegom (2009) simplifies. Here the null hypothesis to be tested for model (\ref{mod})  is
$$\bar{H}_0:\exists c \mbox{ s.\,t.\ } m=c\sigma. $$
This condition connects  ARCH models of the form $Z_t=s(Z_{t-1})\epsilon_t$ to model (\ref{mod}) by setting $X_t=Z_t^2=m(X_{t-1})+\sigma(X_{t-1})\e_t$ with $\e_t=\epsilon_t^2-1$ and $c=(E[\epsilon_1^4]-1)^{-1/2}$.

Note that under $\bar{H}_0$ the constant $c$ can be estimated $\sqrt{n}$-consistently by a least-squares estimator $\hat c$ defined by Dette, Pardo-Fernßndez \& Van Keilegom (2009).
Its asymptotic expansion under $\bar{H}_0$ and under our assumptions is 
\begin{eqnarray}\label{c-ent}
\hat c-c&=&\frac{1}{n}\sum_{t=1}^n\Big(-\frac{1}{2}c\e_t^2+\e_t+\frac{1}{2}c\Big)\frac{\sigma^4(X_{t-1})}{E[\sigma^4(X_0)]}+\op,
\end{eqnarray}
see Theorem 5 in the aforementioned paper, but note that under our assumptions given in the appendix the influence of the weight function vanishes asymptotically. 
 Now define
\begin{eqnarray*}
\hat{F}_{0,n}(y)&=&\sum_{t=1}^nv_{n,t}I\{\hat{\e}_{0,t}\leq y\}
\end{eqnarray*}
with residuals 
$$\hat\e_{0,t} = \frac{X_t-\hat c\hat\sigma(X_{t-1})}{\hat\sigma(X_{t-1})},$$
where $\hat\sigma$ is as in (\ref{m-sig}). Let for $k\in\{4,8\}$,
$$s_k =\sum_{t=1}^n v_{n,t}\hat\sigma^k(X_{t-1})$$
and $\hat\tau^2=s_8/s_4^2-1$. Finally, let $\hat F_n$ be defined as in (\ref{Fn}). Then we have the following result. 

\begin{theo} Under the assumptions stated in the appendix for model (\ref{mod}) with Gaussian innovations we have under $\bar{H}_0$ that 
$$T_n=\frac{2\sqrt{\pi}n\int_\er (\hat F_n(y)-\hat F_{0,n}(y))^2\,dy}{(\frac{3}{4}\hat c^2+1)^2\hat\tau^2}$$
converges in distribution a $\chi^2_1$-distributed random variable.
\end{theo}

{\bf Proof.} 
We only sketch the main differences to the proof in Dette, Pardo-Fernßndez \& Van Keilegom (2009) due to slightly different assumptions (under which in particular the influence of the weight function is asymptotically negligible).  
We have
\begin{eqnarray*}
\hat{F}_{0,n}(y)&=&\sum_{t=1}^nv_{n,t}I\Big\{{\e}_t\leq \frac{(\hat c\hat\sigma-c\sigma)(X_{t-1})}{\sigma(X_{t-1})}+y\frac{\hat\sigma(X_{t-1})}{\sigma(X_{t-1})}\Big\}\\
&=&\frac 1n\sum_{t=1}^nI\{\e_t\leq y\}+\varphi(y)\sum_{t=1}^nv_{n,t}\Big(\frac{(\hat c\hat\sigma-c\sigma)X_{t-1}}{\sigma(X_{t-1})}+y\frac{\hat\sigma^2(X_{t-1})-\sigma^2(X_{t-1})}{2\sigma^2(X_{t-1})}\Big) \\
&&{}+\op
\end{eqnarray*}
uniformly with respect to $y\in\er$ (compare to (\ref{Fn-ent})). From this and (\ref{c-ent}), (\ref{Fn-ent}), (\ref{sig-ent}) one obtains 
$$\sqrt{n}(\hat F_n(y)-\hat F_{0,n}(y))=\varphi(y)\frac{1}{\sqrt{n}}\sum_{t=1}^n 
\Big(\frac{1}{2}c(\e_t^2-1)-\e_t\Big)
\Big(1-\frac{\sigma^4(X_{t-1})}{E[\sigma^4(X_0)]}\Big)+o_P(1)
$$
which by Th.\ 2.21 in Fan \& Yao (2003) converges in distribution to $\varphi(y)Z$, where $Z$ is centered normally distributed with variance $(\frac{3}{4}c^2+1)\tau^2$ with 
$$\tau^2=E\Big[\Big(1-\frac{\sigma^4(X_{t-1})}{E[\sigma^4(X_0)]}\Big)^2\Big]=\frac{E[\sigma^8(X_0)]}{(E[\sigma^4(X_0)])^2}-1.$$ Finally the assertion follows because $\hat\tau^2$ consistently estimates $\tau^2$ because $\hat{\sigma}$ consistently estimates $\sigma$ and $(\sigma(X_t))_{t\in\zet}$ inherits the mixing property of $(X_t)_{t\in\zet}$ and is therefore ergodic as well. Thus
$T_n$ converges in distribution to 
$$2\sqrt{\pi}\int_\er (\varphi(y))^2\,dy \,\frac{Z^2}{(\frac{3}{4}c^2+1)\tau^2}=\frac{Z^2}{(\frac{3}{4}c^2+1)\tau^2}, $$
which is $\chi^2_1$-distributed. 
\hfill $\Box$

\medskip

One obtains an asymptotically distribution-free test for $\bar{H}_0$ and thus avoids to implement bootstrap procedures. Note that it is not obvious which kind of bootstrap procedure should be applied here in the context of arbitrary innovation distributions.

\begin{appendix}

\section{Technical assumptions}\label{section-tech}
\def\theequation{A.\arabic{equation}}
\setcounter{equation}{0}

The assumptions are similar to those in Selk \& Neumeyer (2012) and required for their Theorem 3.1 which we use.

\begin{description}
\item [(K)] The kernel $K$ is a three times differentiable density with compact support $[-C,C]$ and $\sup_{u\in[-C,C]}|K^{(\mu)}(u)|\leq \bK <\infty$, $\mu=0,1,2,3$. Moreover $K(C)=K(-C)=K'(C)=K'(-C)=0$ and $\int K(u)udu=0$.

\item[(C)] The sequence of bandwidths $c_n$ fulfills 
$$nc_n^4(\log n)^\eta\to 0,\quad \frac{(\log n)^\eta}{nc_n^{2+\sqrt{3}}}\to 0\mbox{ for all } \eta>0.$$


\item [(I)] For the interval $I_n=[a_n,b_n]$ some $r_I<\infty$ exists, such that $(b_n-a_n)=O(\log(n)^{r_I})$. Moreover $\left(\int _{-\infty}^{a_n+\kappa}f_{X_0}(x)dx+\int_{b_n-\kappa}^{\infty}f_{X_0}(x)dx\right)=o((\log n)^{-1})$, where $f_{X_0}$ denotes the density of $X_0$.

\item [(W)] The weight function $w_n:\er\to[0,1]$ fulfills $w_n(x)=1$ for $x\in[a_n+\kappa,b_n-\kappa]$ and $w_n(x)=0$ for $x\notin[a_n,b_n]$ for some $\kappa>0$ independent of $n$ and is three times differentiable such that  $\sup_{n\in\en}\sup_{x\in\er}|w_n^{(\mu)}(x)|<\infty$ for $\mu=1,2,3$.

\item [(F)] The innovations $\e_j$, $j\in\mathbb{Z}$, are independent and standard normally distributed.

\item [(E)] Some $b> 1+\sqrt 3$ exists such that $E\left[|X_0|^{2b}\right]<\infty$.

\item [(X)]The observation process $(X_j)_{j\in\zet}$ is $\alpha$-mixing with mixing-coefficient $\alpha(n)=O(n^{-\beta})$ for some \[\beta>\max\left(2\frac{(3+\sqrt 3)b+2+\sqrt 3}{(1+\sqrt 3)b-2(2+\sqrt3)}, 7\right). \]
Their density $f_{X_0}$ is bounded and four times differentiable with bounded derivatives.
The density is also bounded away from zero on compact intervals and some $r_f<\infty$ exists, such that $\frac 1{\inf_{x\in I_n}f_{X_0}(x)}=O((\log n)^{r_f})$.



\item[(Z)]
It holds that $$\sup_{x\in J_n}\left(\left( |m(x)|+|\sigma(x)|\right)^{2k}f_{X_{0}}(x)\right)=O(1)$$ 
 and there exists some $1\leq j^*<\infty$ such that 
 $$\sup_{x,x'\in J_n}\left(\left(|m(x)|+|\sigma(x)|\right)^k\left(|m(x')|+|\sigma(x')|\right)^kf_{X_{0},X_{j-1}}(x,x')\right)=O(1)$$ is valid for all $j>j^*+1$,
 for $k=1,2$, $n\to\infty$ with  $J_n=[a_n-(C+c_n^{-\frac 12}n^{-\frac 12}(\log n)^{\frac 12})c_n\ ,\ b_n+(C+c_n^{-\frac 12}n^{-\frac 12}(\log n)^{\frac 12})c_n]$.

\item [(M)] The regression function $m$ and the scale function  $\sigma$ are four times differentiable and there exist some $r_q,r_s<\infty$ and $q_n$, $q_n^{\sigma}$ with $\sup_{x\in [a_n-Cc_n,b_n+Cc_n]}|m^{(\mu)}(x)|=O(q_n)$,   $\sup_{x\in [a_n-Cc_n,b_n+Cc_n]}|\sigma^{(\mu)}(x)|=O(q_n)$, $\mu=0,1,2,3,4$ and $\frac 1{\inf_{x\in I_n}|\sigma(x)|}=O(q_n^{\sigma})$, where $q_n=O((\log n)^{r_q})$, $q_n^{\sigma}=O((\log n)^{r_s})$, $(q_n)^{-1}=O(1)$, $(q_n^{\sigma})^{-1}=O(1)$.
\end{description}

\begin{rem}\rm 
Note that the mixing condition in (X) is weaker than the one in Selk \& Neumeyer (2012). This is due to the nonsequential case that is examined here for which the proof of Lemma B.3 in the aforementioned paper can be simplified.
Further note that the assumptions above are formulated under the null hypothesis $H_0$ of Gaussian innovations. To obtain consistency of the testing procedures one needs to replace assumption (F) by
\begin{description}
\item [(F')] The innovations $\e_j$, $j\in\mathbb{Z}$, are independent and identically distributed with distribution function $F$. Their density $f$ is continuously differentiable and $\sup_{t\in\er}|f(t)t|<\infty$  as well as $\sup_{t\in\er}|f'(t)t^2|<\infty$. Further, $E\left[|\e_1|^{2b}\right]<\infty$ for $b$ from assumption (E). 
\end{description}
For AR model (\ref{mod-AR}) with $\eta_t=\sigma\eps_t$ some conditions in (Z) and (M) are redundant because $\sigma$ is a constant function. A similar remark holds for the ARCH model (\ref{mod-ARCH}) where $m\equiv 0$. 
$\blacksquare$
\end{rem}

\vspace{1cm}


\section*{References}

\begin{description}
\item Akritas, M. \& Van Keilegom, I. (2001). \textit{Nonparametric estimation of the residual distribution.} Scand. J. Statist. 28, 549-567.

\item Araveeporn, A. (2011). \textit{Developing Nonparametric Conditional Heteroscedastic Autoregressive Nonlinear Model by Using Maximum
Likelihood Method.} Chiang Mai J. Sci. 38, 331-345.

\item Brockwell, P. J. \& Davis, R. A. (2006). \textit{Time Series: Theory and Methods}. Springer, New York.

 \item Dette, H., Pardo-Fernßndez, J. C. \& Van Keilegom, I. (2009). \textit{Goodness-of-Fit Tests for Multiplicative Models with Dependent Data}. Scand. J. Statist. 36, 782-799.
 
 \item Doukhan, P. (1994). \textit{Mixing, Properties and Examples}. Springer, New York.

 \item Doukhan, P. \& GhindÞs, M. (1983). \textit{Estimation de la transition de probabilitÚ d'une cha\^{i}ne de Markov Do\"{e}blin-rÚcurrente. ╔tude du cas du processus autorÚgressif gÚnÚral d'ordre 1.} Stochastic Process. Appl. 15, 271-293.
 
\item Ducharme, G. \& Lafaye de Micheaux, P. (2004). \textit{Goodness-of-fit tests of normality  for the innovations in ARMA models}. J. Time Ser. Anal. 25, 373-395
 
 \item Franke, J., Krei▀, J.-P. \& Mammen, E. (2009). \textit{Nonparametric modelling of financial time series}. T. G. Andersen (ed) et al, Handbook of Financial Time Series. Springer, Berlin, 927-952.

\item Hõrdle, W. \& Tsybakov, A. (1997). \textit{Local polynomial estimators of the volatility function in nonparametric autoregression.} J. Econometrics 81, 223-242.

\item Hõrdle, W. \& Vieu P. (1991). \textit{Kernel regression smoothing of time series.} J. Time Ser. Anal. 13, 209-232.

\item Hong-zhi, A. \& Bing, C. (1991). \textit{A Kolmogorov-Smirnov type statistic with application to test for nonlinearity in time series.} Internat. Statist. Rev. 59, 287-307.

\item Horvßth, L., Kokoszka, P. \& TeyssiÞre, G. (2004). \textit{Bootstrap misspecification tests for ARCH based on the empirical process of squared residuals.} J. Stat. Comput. Simul. 74 , 469-485.

\item Klar, B., Lindner, F. \& Meintanis, S. G. (2011). \textit{Specification tests for the error distribution in GARCH models.} Comput. Statist. Data Anal., to appear. preprint available at http://www.math.kit.edu/stoch/~klar/seite/veroeffentlichungen/de

\item Koul, H. L. \& Ling, S. (2006). \textit{Fitting an error distribution in some heteroscedastic time series models.} Ann. Statist. 34, 994-1012. 

\item Koul, H. L. \& Stute, W. (1999). \textit{Nonparametric model checks for time series.} Ann. Statist. 27, 204-236.

\item Masry, E. \& Tj$\o$stheim, D. (1995). \textit{Nonparametric estimation and identification of nonlinear ARCH time series.} Econometric Theory 11, 258-289.

\item M³ller, U. U., Schick, A. \& Wefelmeyer, W. (2009). \textit{Estimating the innovation distribution in nonparametric autoregression}. Probab. Theory Relat. Fields 144, 53-77.

\item Neumeyer, N. \& Van Keilegom, I. (2010). \textit{Estimating the error distribution in nonparametric multiple regression with applications to model testing.} J. Multiv. Anal. 101, 1067-1078. 

\item Robinson, P. M. (1983). \textit{Nonparametric estimators for time series.} J. Time Ser. Anal. 4, 185-207.

\item Selk, L. \& Neumeyer, N. (2012). \textit{Testing for a change of the innovation distribution in nonparametric autoregression - the sequential empirical process approach}. Preprint available at http://preprint.math.uni-hamburg.de/public/ims.html

\item Shorack, G. R. \&  Wellner, J.\,A. (1986). \textit{Empirical Processes with Applications ot Statistics.} Wiley, New York. 

\item Stephens, M. A. (1976). \textit{Asymptotic Results for Goodness-of-Fit Statistics with Unknown Parameters}. Ann. Statist. 4, 357-369.

\item Van Keilegom, I., Gonzßlez Manteiga, W. \& Sßnchez Sellero, C. (2008). \textit{Goodness-of-fit tests in parametric regression based on the estimation of the error distribution.} TEST, 17, 401-415. 
\end{description}

\end{appendix}

\end{document}